\documentclass[sigconf]{acmart}

\AtBeginDocument{%
  \providecommand\BibTeX{{%
    \normalfont B\kern-0.5em{\scshape i\kern-0.25em b}\kern-0.8em\TeX}}}

\usepackage{graphicx}
\usepackage{url}
\usepackage{amsmath}
\usepackage{amsfonts}
\usepackage{color}
\usepackage{subfig}
\usepackage{boldline}
\usepackage{tablefootnote}

\graphicspath{{Images/}}

\newcommand{\puttitle}{Self-Imitation Learning of Locomotion Movements through Termination Curriculum}
\newcommand{\putvideo}{supplemental video\footnote{\url{https://youtu.be/dxTZk35Ofyg}}}
\newcommand{\putsourcecode}{\url{https://github.com/donamin/self-imitation-learning}}

\setcopyright{acmcopyright}

\begin{document}

\title{\puttitle}


\author{Amin Babadi}
\email{amin.babadi@aalto.fi}
\affiliation{%
  \institution{Department of Computer Science\\Aalto University}
  \city{Helsinki}
  \state{Finland}
}

\author{Kourosh Naderi}
\email{kourosh.naderi@aalto.fi}
\affiliation{%
  \institution{Department of Computer Science\\Aalto University}
  \city{Helsinki}
  \state{Finland}
}

\author{Perttu H\"{a}m\"{a}l\"{a}inen}
\email{perttu.hamalainen@aalto.fi}
\affiliation{%
  \institution{Department of Computer Science\\Aalto University}
  \city{Helsinki}
  \state{Finland}
}

\renewcommand{\shortauthors}{Babadi et al.}

\begin{abstract}
Animation and machine learning research have shown great advancements in the past decade, leading to robust and powerful methods for learning complex physically-based animations. However, learning can take hours or days, especially if no reference movement data is available. In this paper, we propose and evaluate a novel combination of techniques for accelerating the learning of stable locomotion movements through self-imitation learning of synthetic animations. First, we produce synthetic and cyclic reference movement using a recent online tree search approach that can discover stable walking gaits in a few minutes. This allows us to use reinforcement learning with Reference State Initialization (RSI) to find a neural network controller for imitating the synthesized reference motion. We further accelerate the learning using a novel curriculum learning approach called Termination Curriculum (TC), that adapts the episode termination threshold over time. The combination of the RSI and TC ensures that simulation budget is not wasted in regions of the state space not visited by the final policy. As a result, our agents can learn locomotion skills in just a few hours on a modest 4-core computer. We demonstrate this by producing locomotion movements for a variety of characters.
\end{abstract}


\copyrightyear{2019}
\acmYear{2019}
\acmConference[MIG '19]{Motion, Interaction and Games}{October 28--30, 2019}{Newcastle upon Tyne, United Kingdom}
\acmBooktitle{Motion, Interaction and Games (MIG '19), October 28--30, 2019, Newcastle upon Tyne, United Kingdom}
\acmPrice{15.00}
\acmDOI{10.1145/3359566.3360072}
\acmISBN{978-1-4503-6994-7/19/10}




\maketitle


\section{Introduction}\label{sec:introduction}
Intelligent control of physics simulation is an increasingly popular approach for synthesizing physically-plausible animations for simulated characters. This requires a method that outputs, for each timestep, simulation actuation parameters such as joint torques such that the character performs some desired movement. This poses a continuous control problem with high state and control dimensionality, where the environment is governed by complex physical interactions.

Current approaches for solving physically-based animation can be divided into two categories: 1) planning and search, and 2) reinforcement learning. Planning and search methods use the interleaved mechanism of iterative generation and evaluation of candidate solutions until finding a sufficiently good one. On the other hand, reinforcement learning (RL) methods learn how to act through interaction with the environment.

In this paper, we propose a self-imitation learning approach for enabling rapid learning of stable locomotion controllers. Essentially, our approach combines FDI-MCTS \citep{Rajamaeki2018} and DeepMimic \citep{2018-TOG-deepMimic}, two recent methods for continuous control. We are motivated to mitigate the main limitations of both methods, namely the high run-time cost of FDI-MCTS and the data-dependency and sample complexity of DeepMimic.

We begin by using FDI-MCTS to generate a cyclic locomotion movement as the reference motion. Then we employ a training mechanism similar to DeepMimic, to find a neural network controller that imitates the reference motion. We also propose Termination Curriculum (TC), a novel curriculum learning approach for accelerating the imitation learning. Our experiments show that our approach is able to learn robust locomotion skills for a broad set of 3D characters. All controllers are trained in less than four hours, which is significantly faster than DeepMimic and SFV.


\section{Related Work}\label{Sec-Related}
Our work aims at producing locomotion movements for physically-based characters. There has been a large body of research on this problem, especially after remarkable breakthroughs in Deep Reinforcement Learning (DRL) \citep{mnih2015human, silver2016mastering, silver2017mastering}. Proposed approaches can be divided into two categories: planning and search (Section \ref{Sec-Related-PlanningAndSearch}), and reinforcement learning (Section \ref{Sec-Related-RL}).

\subsection{Planning and Search}\label{Sec-Related-PlanningAndSearch}
Evolutionary Strategies (ES) are a family of black-box optimization methods that are also very easy to use in parallel \citep{salimans2017evolution}. One of the most common ES methods is Covariance Matrix Adaptation Evolution Strategy (CMA-ES) \citep{hansen2006cma}. CMA-ES has been used in an offline manner to learn the parameters for controller of physically-simulated characters \citep{Geijtenbeek2013}. Another study has used CMA-ES as an offline low-level controller for synthesizing humanoid wall climbing movements \citep{naderi2017discovering}. Recent studies have used CMA-ES for synthesizing sports movements in a two-player martial arts interface \citep{babadi2018intelligent} and single-agent basketball \citep{liu2018learning}.

Monte Carlo methods have received considerable interest in domains where the search budget is limited. Sequential Monte Carlo (SMC) has been shown to be effective for online synthesis of physically-based animations \citep{hamalainen2014online}. It has also been used for replicating motion capture data by breaking the problem into a sequence of control fragments \citep{liu2016guided}. Another Monte Carlo method, called Monte Carlo Tree Search (MCTS) \citep{browne2012survey}, has shown great performance in real-time applications and games \citep{sironi2018self}. MCTS has been used in Alpha Go Zero for improving the policy using self-play \citep{silver2017mastering}. Fixed Depth Informed MCTS (FDI-MCTS) is a continuous version of MCTS used in physically-based control. FDI-MCTS uses a policy network trained with supervised learning to reduce the movement noise caused by the sampling-based controller \citep{Rajamaeki2018}.

\subsection{Reinforcement Learning}\label{Sec-Related-RL}
Reinforcement Learning (RL) methods have become significantly more powerful in the recent years, mainly after the Deep Reinforcement Learning (DRL) success in Atari games \citep{mnih2015human}. Two of the most common RL algorithms are called Trust Region Policy Optimization (TRPO) \citep{schulman2015trust} and Proximal Policy Optimization (PPO) \citep{schulman2017proximal}. The key element in these methods is a surrogate objective function that allows for more than one gradient update per data sample. Studies have shown that PPO outperforms TRPO in most cases, which makes it the dominant algorithm used in continuous control \citep{yu2018learning, 2018-TOG-deepMimic}.

Curriculum learning is the process of learning a series of tasks in increasing order of complexity \citep{bengio2009curriculum}. It is a powerful technique for improving learning performance in terms of both convergence speed and output quality. This technique has been used for learning humanoid climbing movements, where the agent learns 1-, 2-, 3-, and 4-limb movements in order (a limb can be either one of agent's hands or feet) \citep{naderi2018learning}. A recent study proposed a continuous curriculum learning method for providing physical assistance to help the character in locomotion movements \citep{yu2018learning}.

A popular data-driven approach for solving physically-based control is to learn through imitation. An example in this category is called DeepLoco, a system that trains high-level and low-level controllers such that the low-level controller is encouraged to imitate a reference animation \citep{peng2017deeploco}. A descendant of this method, called DeepMimic, has been able to produce a wide set of high-quality and robust movements by imitating a large motion capture dataset \citep{2018-TOG-deepMimic}. The most recent variant of this method is able to extract reference animations directly from videos, which makes the training pipeline significantly cheaper \citep{2018-TOG-SFV}.

\section{Preliminaries}\label{Sec-Background}

Online optimization is one of the main approaches for generating physically-based animations \citep{geijtenbeek2012interactive}. The idea is to generate a set of candidate solutions (i.e., action sequences), and find the most cost-efficient solution by evaluating them. This is usually done using forward simulation until some time horizon and computing some cost function that encodes some information about the target movement. In the walking task for example, the cost function encourages the character to keep its center of mass above its feet and its mean velocity close to some desired walking velocity. The function also usually penalizes the amount of torque applied to each body joint in order to avoid unrealistic movements \citep{rajamaki2017augmenting}.

We use a recent open-source\footnote{\url{https://github.com/JooseRajamaeki/TVCG18}} online optimization method called \textit{Fixed-Depth Informed Monte Carlo Tree Search (FDI-MCTS)} \citep{Rajamaeki2018} to produce a cyclic locomotion movement as the reference motion. FDI-MCTS synthesizes movements using an interleaved process of tree search and supervised learning. The supervised learning component is trained using the best controls found during the tree search in order to reduce the noise in produced movements. This results in emergence of stable locomotion gaits for different types of characters in less than a minute, allowing rapid cost function design iteration. Movement is initially noisy, but running the algorithm for a few more minutes removes the noise.

The basic definition of reinforcement learning includes an agent that has interactions with some environment, and its goal is to maximize the accumulated rewards over time \citep{sutton2018reinforcement}. Policy optimization refers to a family of reinforcement learning methods, in which the goal is to optimize the agent's policy with respect to the expected return. The policy is usually modelled using a neural network, and defines a mapping from a state to a distribution over actions.

In this paper, we use an open-source\footnote{\url{https://github.com/openai/baselines}} implementation of \textit{Proximal Policy Optimization (PPO)} \citep{schulman2017proximal}. PPO uses stochastic gradient ascent by estimating the gradient of the expected return with respect to the policy parameters. It does that using a so-called \textit{clipped surrogate objective function}, that penalizes large policy updates.

\section{Method}\label{Sec-Method}


\begin{figure*}[!ht]
\centering
\subfloat[Wolf]{{\includegraphics[width=0.33\linewidth]{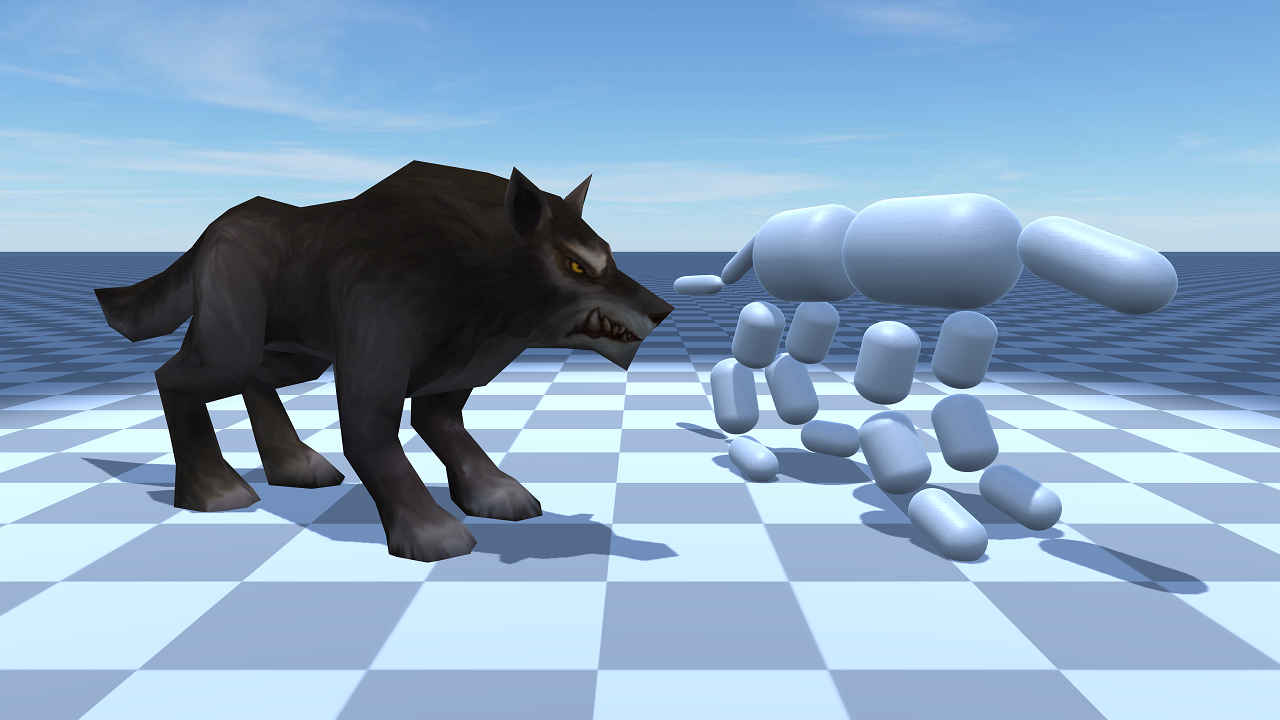} }\label{Fig:rigs-wolf}}
\subfloat[Orc]
{{\includegraphics[width=0.33\linewidth]{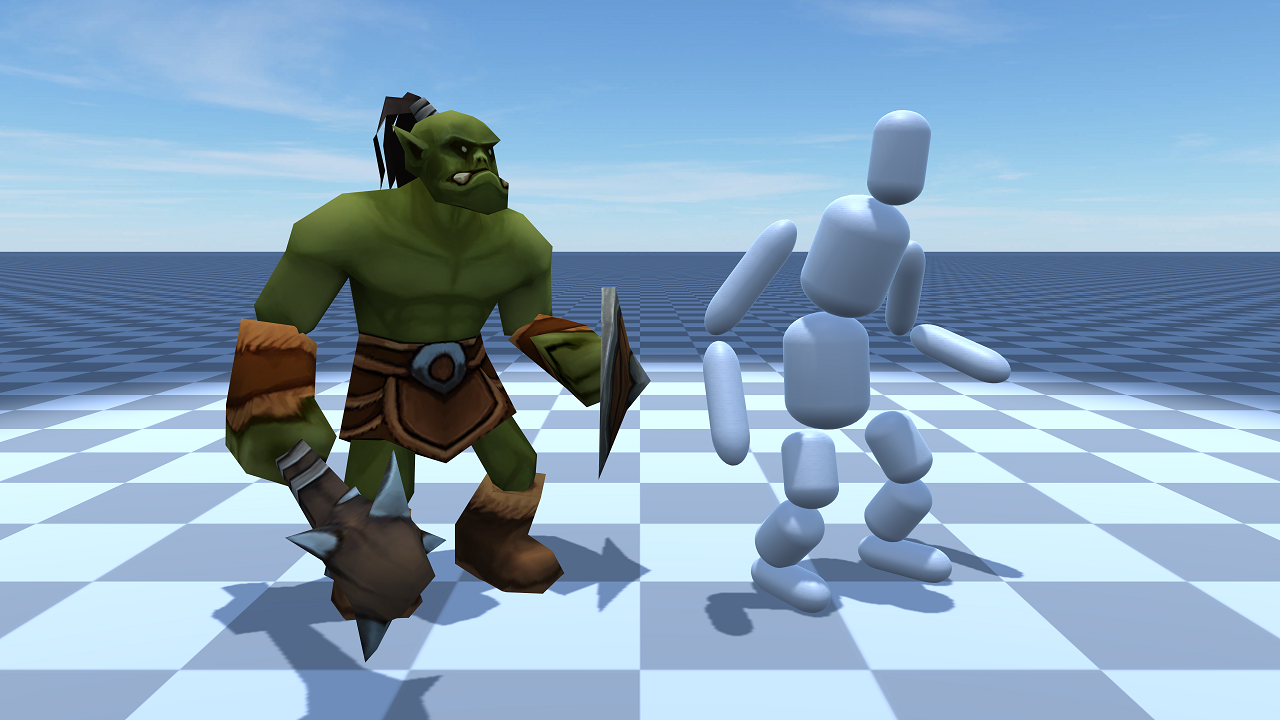} }\label{Fig:rigs-orc}}
\subfloat[Mech]{{\includegraphics[width=0.33\linewidth]{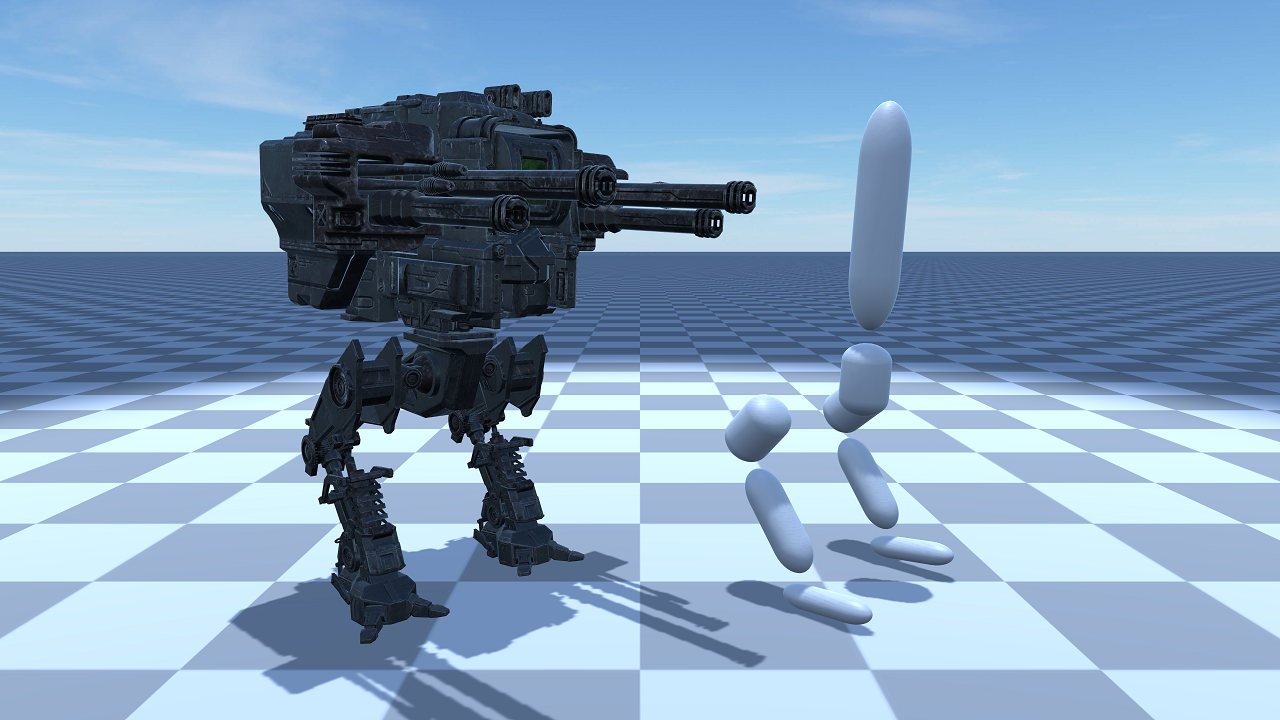} }\label{Fig:rigs-mech}}
\caption{Simulated 3D characters and their physical skeletons in the default pose. Since our approach does not require any hand-designed or mocap animations, it can be applied to a wide range of character anatomies with minimum cost.}
\label{Fig:rigs}
\end{figure*}

\subsection{Reference Motion Generation}
The first step of our approach includes automatic generation of a reference motion. For this purpose, we use FDI-MCTS \citep{Rajamaeki2018}, a recent sampling-based model-predictive algorithm for continuous control. The cost function used by FDI-MCTS uses four quadratic terms for penalizing the followings: 1) amount of torques applied to joints, 2) deviation from the default pose (shown in Fig. \ref{Fig:rigs}), 3) planar deviation of the center of mass from the feet mean point, and 4) the difference between the current and the target velocity of the character. When the locomotion gait has become stable (after a few minutes in our experiments), we extract a cycle out of the synthesized motion sequence using the method explained next.

Given a trajectory of locomotion movements, the cycle extraction process starts by storing the key information at every timestep. Stored information include orientation $q^j_t$ of each joint $j$, angular velocity $\dot{q}^b_t$ of each bone $b$, position $p^e_t$ and linear velocity $v^e_t$ of each end-effector $e$ (e.g., hands and feet for humanoid characters), and the character's center of mass $p^{com}_t$. At each timestep, in order to detect the end of the cycle, positions and linear velocities of all end-effectors are compared with their corresponding values at the initial timestep of the cycle. The cycle is completed if the end-effectors have almost the same positions as in the initial timestep and their corresponding linear velocities have acute angles between them, i.e., their dot product is positive. We set the minimum length of a cycle to $10$ timesteps to avoid detecting empty cycles. This gives us an easy-to-implement and computationally cheap method for extracting cycles from synthesized movements.

\subsection{Self-Imitation Learning}

After synthesizing a cyclic movement, we use PPO to find a policy for performing locomotion while imitating the reference motion. In this part, we use a training mechanism similar to DeepMimic \citep{2018-TOG-deepMimic}. When starting a new episode, the so-called \textit{Reference State Initialization (RSI)} is used, i.e., the initial state is uniformly picked from the reference motion. An episode is terminated if a bone other that the feet is in contact with the ground, or episode length exceeds some pre-defined limit. To accelerate the training process, we employ a different \textit{Early Termination (ET)} mechanism, which is explained in Section \ref{SubSection:ContinuousCurriculumLearning}.

\subsection{Reward Function}\label{subsec-reward-function} Our reward definition is almost identical to DeepMimic \citep{2018-TOG-deepMimic}. The instantaneous reward $r_t$ at timestep $t$ is defined as follows:

$$r_t=\omega^I r^I_t + \omega^T r^T_t,$$
where $r^I$ (weighted by $\omega^I$) and $r^T$ (weighted by $\omega^T$) define the imitation and task rewards at timestep $t$. This encourages the character to satisfy the task objective while imitating the reference motion. All quantities on the right side of the equation are between $0$ and $1$, and $\omega^I+\omega^T=1$, leading to a bounded reward $0 \leq r_t \leq 1$.

\subsubsection{Imitation Reward} The imitation reward encourages the character to imitate the reference motion, and it is computed as a weighted sum of four terms as follows:

$$
r^I_t = w^p r^p_t+w^v r^v_t+w^e r^e_t+w^c r^c_t
$$

$$
w^p=0.65, w^v=0.1, w^e=0.15, w^c=0.1
$$

The terms $r^p$, $r^v$, $r^e$, and $r^c$ are computed exactly the same as in \citep{2018-TOG-deepMimic}. 

\subsubsection{Task Reward} In this paper, we only consider the locomotion tasks because FDI-MCTS is mainly designed to produce locomotion movements. So the task reward encourages the character to walk in the desired direction with the desired speed, and it is defined as follows:

$$
r^T_t = \exp \left[
-2.5 \times
{\Vert v^* - \bar{v}\Vert
}^2
\right],
$$
where $v^*$ is the desired velocity and $\bar{v}$ is the mean velocity of character's bones, projected on the \textit{xy}-plane.

\subsection{Termination Curriculum} \label{SubSection:ContinuousCurriculumLearning}

\begin{figure*}[!ht]
\centering
\subfloat[Fixed state initialization]
{{\includegraphics[width=0.24\linewidth]{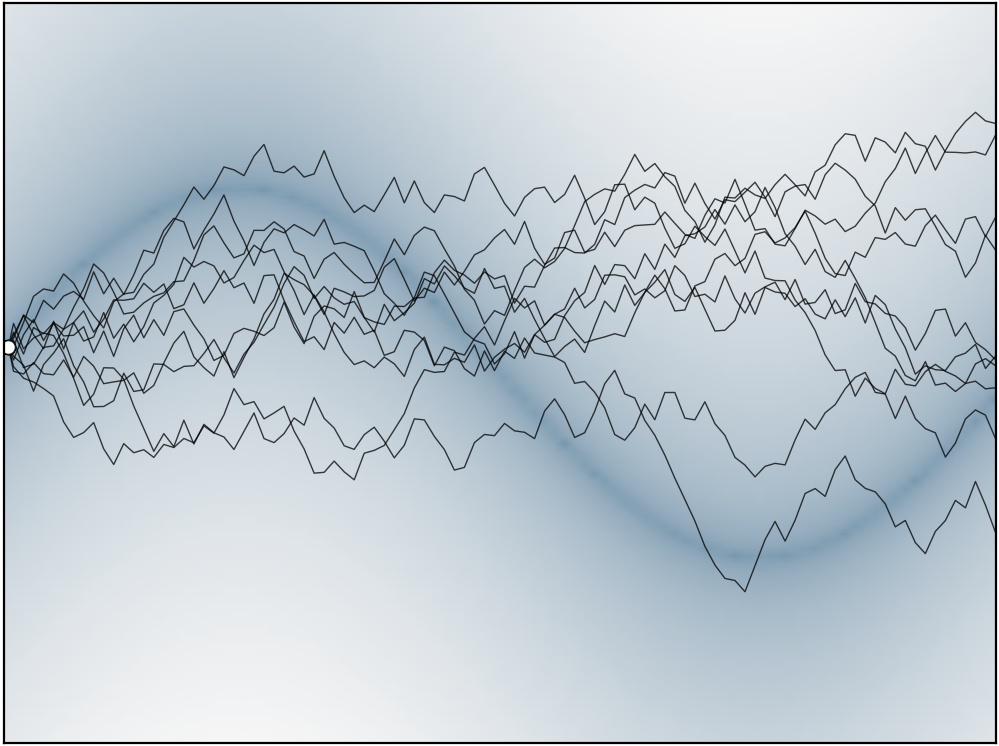} }\label{Fig:toy_problem-A}}
\hfill
\subfloat[Reference state initialization]{{\includegraphics[width=0.24\linewidth]{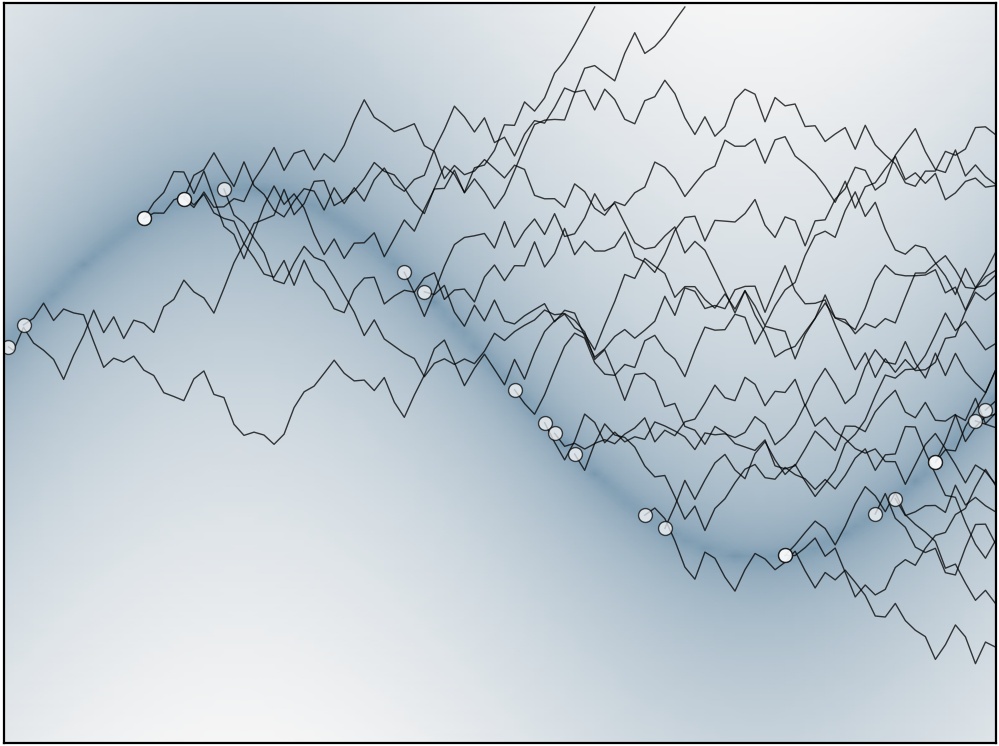} }\label{Fig:toy_problem-B}}
\hfill
\subfloat[Reference state initialization with a high reward threshold in the beginning of training]{{\includegraphics[width=0.24\linewidth]{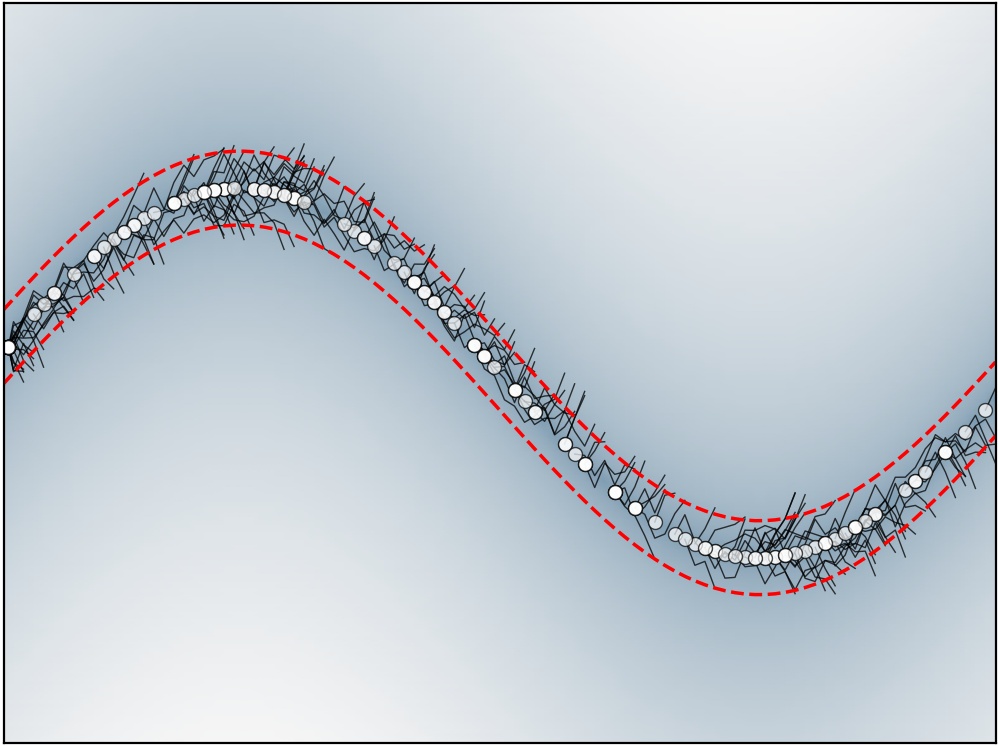} }\label{Fig:toy_problem-C}}
\hfill
\subfloat[Reference state initialization with a high reward threshold in the middle of training]{{\includegraphics[width=0.24\linewidth]{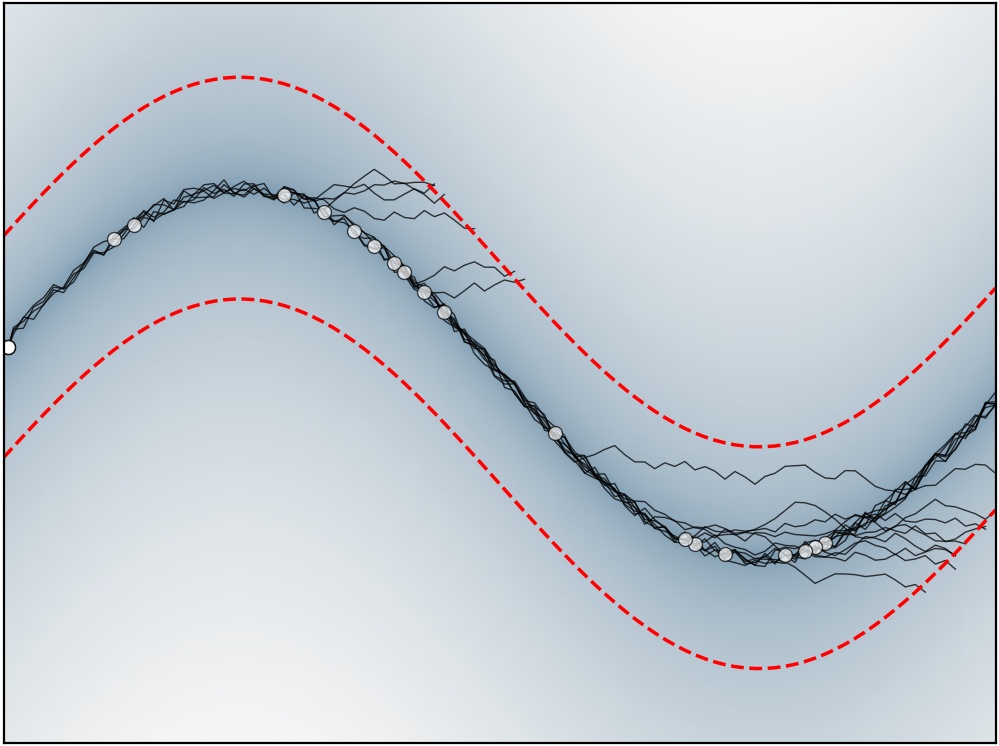} }\label{Fig:toy_problem-D}}
\caption{This didactic example demonstrates the movement of an agent along the vertical axis over time (the horizontal axis). The reward landscape is shown using a light-to-dark heatmap, i.e., the optimal state trajectory is the dark sine curve. The solid black lines are the observed trajectories, whose initial states are shown by white circles. Fixed state initialization can slow down the training process since a slight deviation from the optimal trajectory can easily prevent the agent from visiting promising regions of the state space (a). \textit{Reference State Initialization (RSI)} \citep{2018-TOG-deepMimic} is an effective strategy for mitigating this issue (b). We propose termination curriculum (TC), an early termination strategy, by putting a threshold (shown by the red dashed lines) on the amount of instantaneous reward at each timestep (c). Using this strategy, at early stages of the training the agent will end up observing a lot of short trajectories since the policy is very likely to deviate from the optimal trajectory. However, as the training goes on, the agent will learn to stay in proximity of the optimal trajectory for a longer time, making it possible to reduce the reward threshold (d).}
\label{Fig:toy_problem}
\end{figure*}

When using RL algorithms, a common challenge appears in the initial stages of the training, where the policy can easily lead the agent to the fruitless regions of the state space. Plus, it has been recently shown that PPO is prone to getting stuck in local optima \citep{hamalainen2018ppo}. These can cause a huge waste of simulation budget during the training. We use a didactic example, shown in Fig. \ref{Fig:toy_problem} to demonstrate this problem and how we propose to solve it.

Fig. \ref{Fig:toy_problem-B}, shows how the simulation budget can be wasted even when \textit{Reference State Initialization (RSI)} is used without any termination mechanisms. The figure shows a 2D state space, in which the optimal state trajectory (i.e., the reference motion) is shown in dark gray. The light-to-dark heatmap shows the state-dependent reward distribution and each solid black line resembles a random trajectory (i.e., episode) in the state space. As it can be seen in the figure, RSI forces the episodes to start on the optimal state trajectory. However, a non-optimal policy can easily lead the agent to the regions of the state space that should not be visited using an optimal policy.

To mitigate the problem shown in Fig. \ref{Fig:toy_problem-B}, we propose \textit{Termination Curriculum (TC)}. TC is a continuous curriculum learning mechanism that limits the minimum allowed amount of instantaneous reward in each timestep, and then lowers this limit during training. This simple mechanism can lead to a significant increase of performance in terms of both reward and training speed. Next we explain how TC works in detail.

Let $R^{min}_t$ denote the threshold for the instantaneous reward at timestep $t$. We add an extra termination condition to the underlying MDP such that an episode is terminated if $r_t < R^{min}_t$. This forces the agent to only visit the regions of the state space in which $r_t \geq R^{min}_t$, except in the last timestep of each episode. By lowering the threshold $R^{min}_t$ during the training process, the agent is gradually allowed to visit other regions of the state space as well. In other words, the character performs in a restricted state space, which in the beginning is significantly smaller than the original space. By gradually relaxing the restrictions, the state space becomes larger, allowing the character to learn how to act in the new states.

Fig. \ref{Fig:toy_problem-C} demonstrates how termination curriculum mechanism helps the agent to cover the states that are in proximity to the optimal state trajectory. In this case, the trajectories are shorter. However, when used along with RSI, most of the simulation budget is spent for revising the policy in proximity of the optimal state trajectory. Furthermore, lowering the reward threshold allows the agent to visit more challenging regions of the state space, extending the policy to act in longer episodes.

The main challenge when applying termination curriculum is how to choose the right range for $R^{min}_t$. As explained in Section \ref{subsec-reward-function}, our DeepMimic-style reward function is human-understandable enough, and it is always between $0$ and $1$. Our tests show that starting with $R^{min}_t = 0.75$ and linearly decreasing it to $R^{min}_t = 0.5$ produces good results for a wide range of characters.

\section{Evaluation}\label{Sec-Evaluations}

\subsection{Implementation Details}
We now explain the implementation details for producing the results. The source code is available on GitHub\footnote{\putsourcecode}, and examples of synthesized motions can be seen in the \putvideo.

\paragraph{Physical Simulations:} We used Open Dynamics Engine (ODE) \citep{ode} for physical simulations. Simulations were done in $64$ parallel threads in order to accelerate the optimization and training processes.

\paragraph{State Features:} At each timestep $t$, the state features contain position and orientation of the root bone, angular velocity of each bone, and finally, all the joint angles. These values are concatenated into a vector and used as the feature vector $s_t$.

\paragraph{Action Parameterization:} Action parameters are defined as the reference angles $\alpha_{ref}$ for each degree of freedom. A P-controller then converts these values to reference angular velocities, i.e., $\omega_{ref}=K_P\left(\alpha_{ref} - \alpha_{cur}\right)$, where $K_P=10$ is the P-controller's multiplier and $\alpha_{cur}$ is the current angle.

\paragraph{Training:} We used Tensorflow \citep{tensorflow2015-whitepaper} for training the agent using the PPO algorithm \citep{schulman2017proximal}. Policy and value functions were modeled using fully-connected networks with two hidden layers of size $64$ with $\tanh$ activation function. Our tests showed that it is better to use $\tanh$ in the final layer of the policy network and then interpolate the policy's output using the minimum and maximum angle for each degree of freedom.

\paragraph{Training Parameters:}
The parameters used for PPO training are shown in Table \ref{tbl-trainingparams}.

\begin{table}[h!]
\centering
\caption{Parameters used in the PPO training}
\begin{tabular}{|c|c|}
\hline
\textbf{Parameter}	& \textbf{Value} \\ \hlineB{3.5}
Clipping coefficient $\left(\epsilon\right)$&$0.2$\\ \hline
Number of epochs per iteration & $25$\\ \hline
Learning rate* & $\left[10^{-4}, 10^{-7}\right]$ \\ \hline
Training iterations & $3500$ \\ \hline
Iteration simulation budget & $4096$ \\ \hline
Batch size & $256$ \\ \hline
Discount factor $\left(\gamma\right)$ & $0.99$ \\ \hline
GAE parameter $\left(\lambda\right)$ & $0.95$ \\ \hline
\end{tabular}
\vfill
* Learning rate decay was used throughout the training.
\label{tbl-trainingparams}
\end{table}

\subsection{Experiments}

\subsubsection{Characters}
In order to show the flexibility of our approach, we used a variety of game characters in our simulations\footnote{All 3D models are royalty-free assets purchased from Unity Asset Store.}. The character's physical skeletons were modeled using 3D capsules connected via 3-DOF ball-and-socket or 1-DOF hinge joints. Finally, characters were rendered using Unity 3D game engine\footnote{\url{https://www.unity3d.com/}}. Fig. \ref{Fig:rigs} shows all characters used in our work along with their skeletons. The details of these characters in the physical simulations are shown in Table \ref{TBL:characters_details}.

\begin{table}[htbp]
\caption{Setup details of the simulated characters}
\begin{center}
\begin{tabular}{|c|c|c|c|}
\hline
\textbf{Property}&\textbf{Wolf}&\textbf{Orc}&\textbf{Mech}\\
\hlineB{3.5}
\textbf{Height (m)}&$1.3$&$1.6$&$2.1$\\
\hline
\textbf{Mass (kg)}&$50$&$30$&$30$\\
\hline
\textbf{Bones}&$17$&$13$&$8$\\
\hline
\textbf{Joints}&$16$&$12$&$7$\\
\hline
\textbf{State Dimensions}&$87$&$69$&$37$\\
\hline
\textbf{Action Dimensions (DoF)}&$30$&$24$&$7$\\
\hline
\end{tabular}
\label{TBL:characters_details}
\end{center}
\end{table}

\subsubsection{Experiments}\label{sec:experiments}
We used all characters shown in Fig. \ref{Fig:rigs} for solving the locomotion task, i.e., moving in the forward direction with the target speed of $1m/s$.

\paragraph{Reference motions:} The walking cycles produced in the reference motion generation stage had usually between $20$ to $40$ frames (depending on the character). Thus, the maximum episode length during the training was set to $100$ to ensure that each episode gives the characters enough time for repeating the reference motion at least twice.

\paragraph{Termination curriculum:} In order to show the effectiveness of the termination curriculum mechanism, we tested five different termination strategies as follows (all versions use reference state initialization):

\begin{enumerate}
\item \textbf{No termination:} This version does not use any termination strategies (similar to Fig. \ref{Fig:toy_problem-B}).
\item \textbf{Termination curriculum:} In this version, the training begins by setting the threshold $R^{min}_t$ (introduced in Section \ref{SubSection:ContinuousCurriculumLearning}) to $0.75$ and then linearly decaying it to $0.5$ throughout the training. The next three versions are solely introduced to demonstrate the effectiveness of decaying the threshold and thus use constant values for the threshold $R^{min}_t$.
\item \textbf{Tight threshold:} This version uses the constant threshold $R^{min}_t = 0.75$ for episode termination. This tight threshold only allows the agent to visit the states that are "almost perfect" (similar to Fig. \ref{Fig:toy_problem-C}).
\item \textbf{Medium threshold:} In this version, the constant threshold $R^{min}_t = 0.5$ is used to limit the visible states to those with a fairly good reward value (similar to Fig. \ref{Fig:toy_problem-D}).
\item \textbf{Loose threshold:} Finally, a version with the constant threshold $R^{min}_t = 0.25$ was defined. This threshold does not allow the agent to visit states with very bad rewards, but does not guarantee to keep it in good states.
\end{enumerate}

\paragraph{Reward validation:} In order to show the effectiveness of the reward function defined in Section \ref{subsec-reward-function}, we also trained the agents directly using the FDI-MCTS cost function $\mathcal{J}_t$ converted to reward as:

$$r_t=\frac{b-\mathcal{J}_t}{b},$$
where $b=10000$ is a constant used to map the cost values into the range $[0, 1]$. Note that the reward function gradient is proportional to FDI-MCTS objective function, thereby preserving the online optimization landscape.

\subsection{Results}
Each one of the five versions introduced in Section \ref{sec:experiments} was tested in five independent runs, and the mean and standard deviation of average cost and reward were recorded. In each run, a new walking cycle was generated using FDI-MCTS and then trained using PPO algorithm for $5000$ iterations. All experiments were performed by an Intel Core i7-4930K 3.40GHz CPU processor and 16GB of RAM. Examples of synthesized locomotion movements can be seen in the \putvideo.

\begin{figure*}[!htbp]
\centering
{{\includegraphics[width=0.9\linewidth]{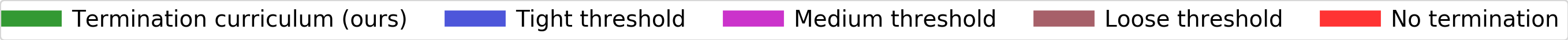} }}
\vfill
\subfloat[Wolf]{{\includegraphics[width=0.33\linewidth]{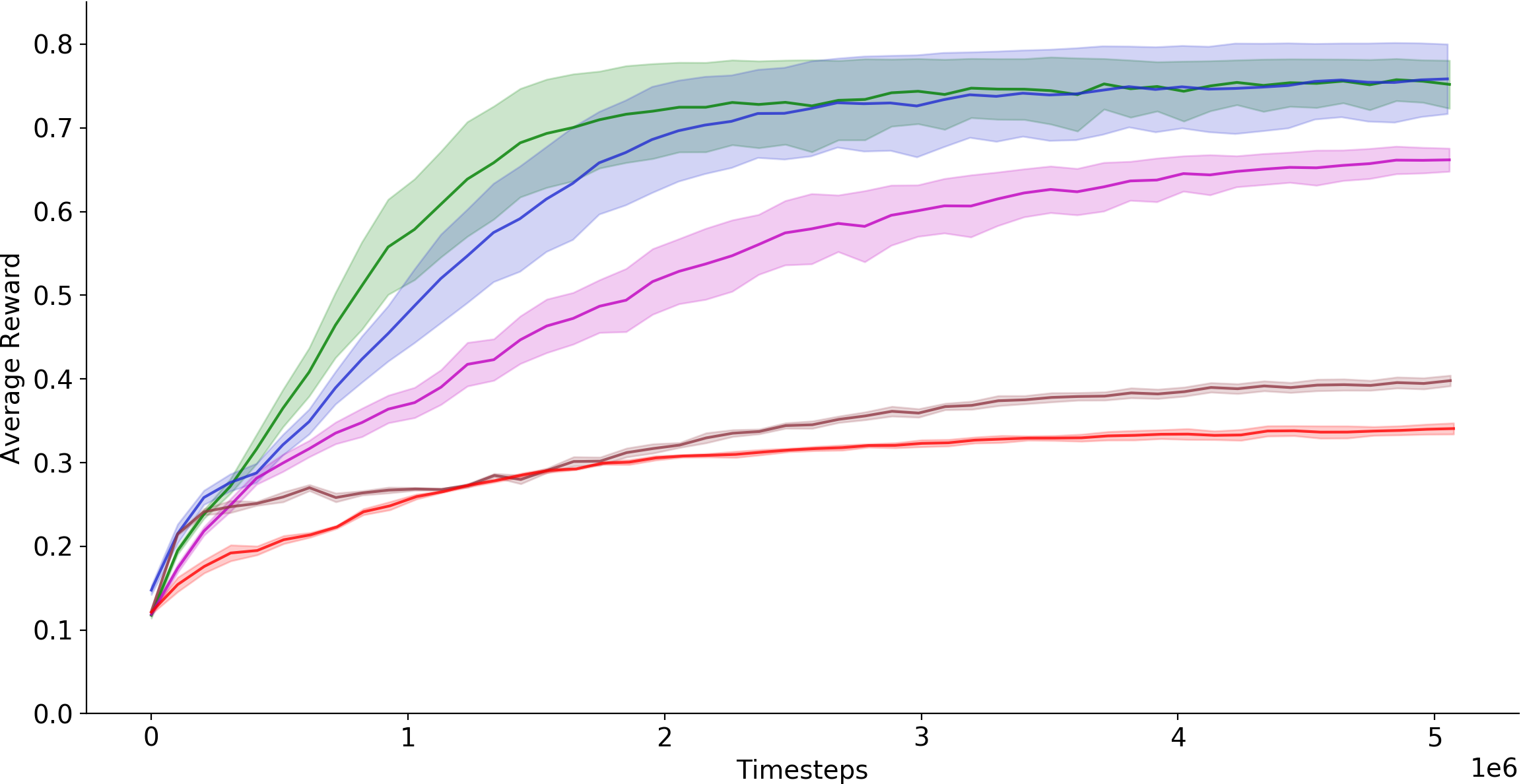} }\label{Fig:reward-wolf}}
\subfloat[Orc]
{{\includegraphics[width=0.33\linewidth]{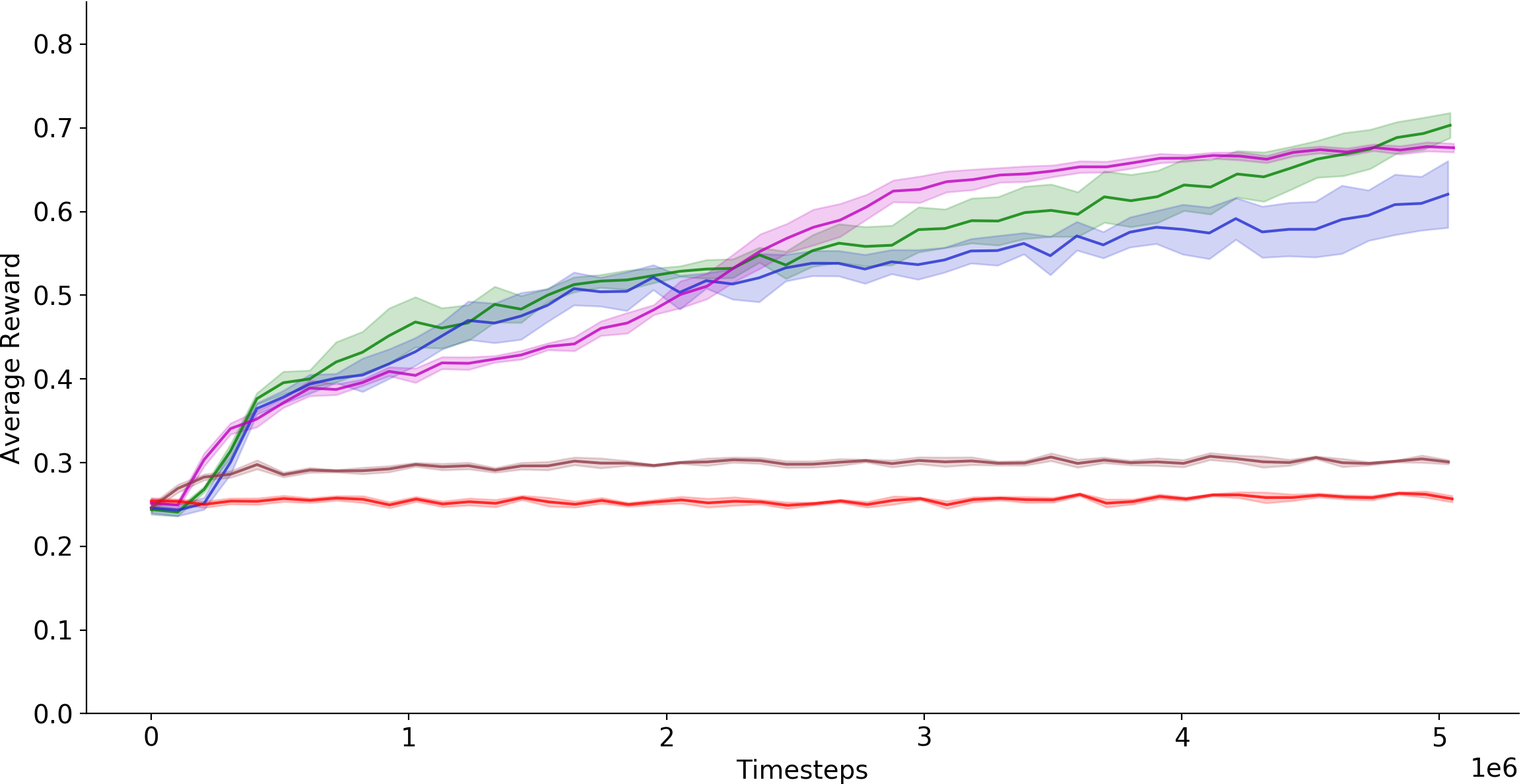} }\label{Fig:reward-orc}}
\subfloat[Mech]{{\includegraphics[width=0.33\linewidth]{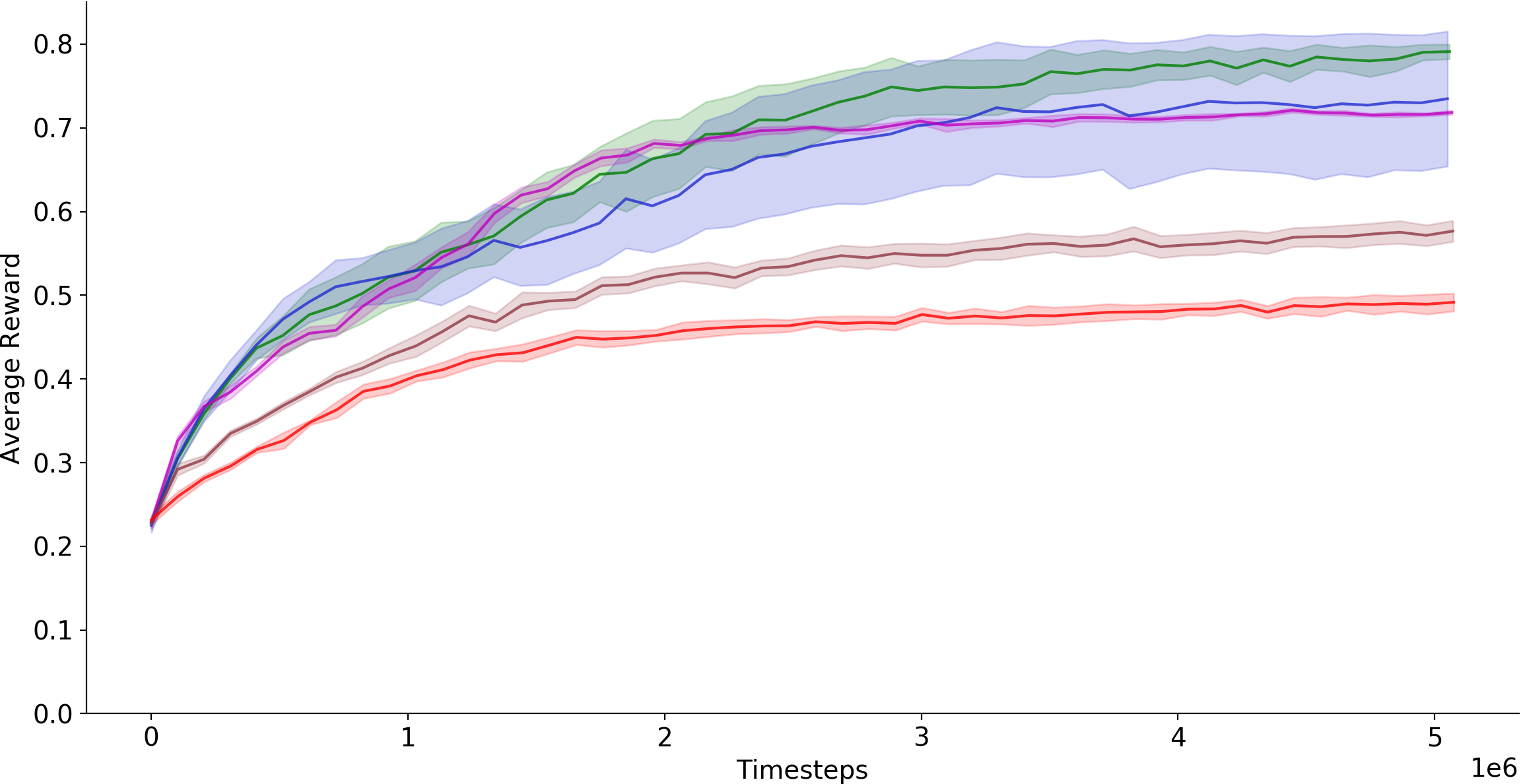} }\label{Fig:reward-mech}}
\caption{Evaluating different termination strategies. Termination Curriculum (TC) significantly accelerates the training, allowing to train characters with various anatomies in only a few hours.}
\label{fig:reward_plot}
\end{figure*}

\begin{figure*}[!htbp]
\centering
{{\includegraphics[width=0.75\linewidth]{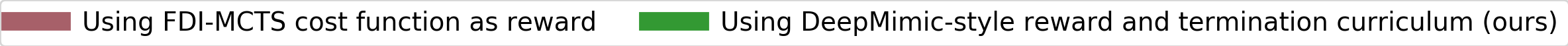} }}
\vfill
\subfloat[Wolf]{{\includegraphics[width=0.33\linewidth]{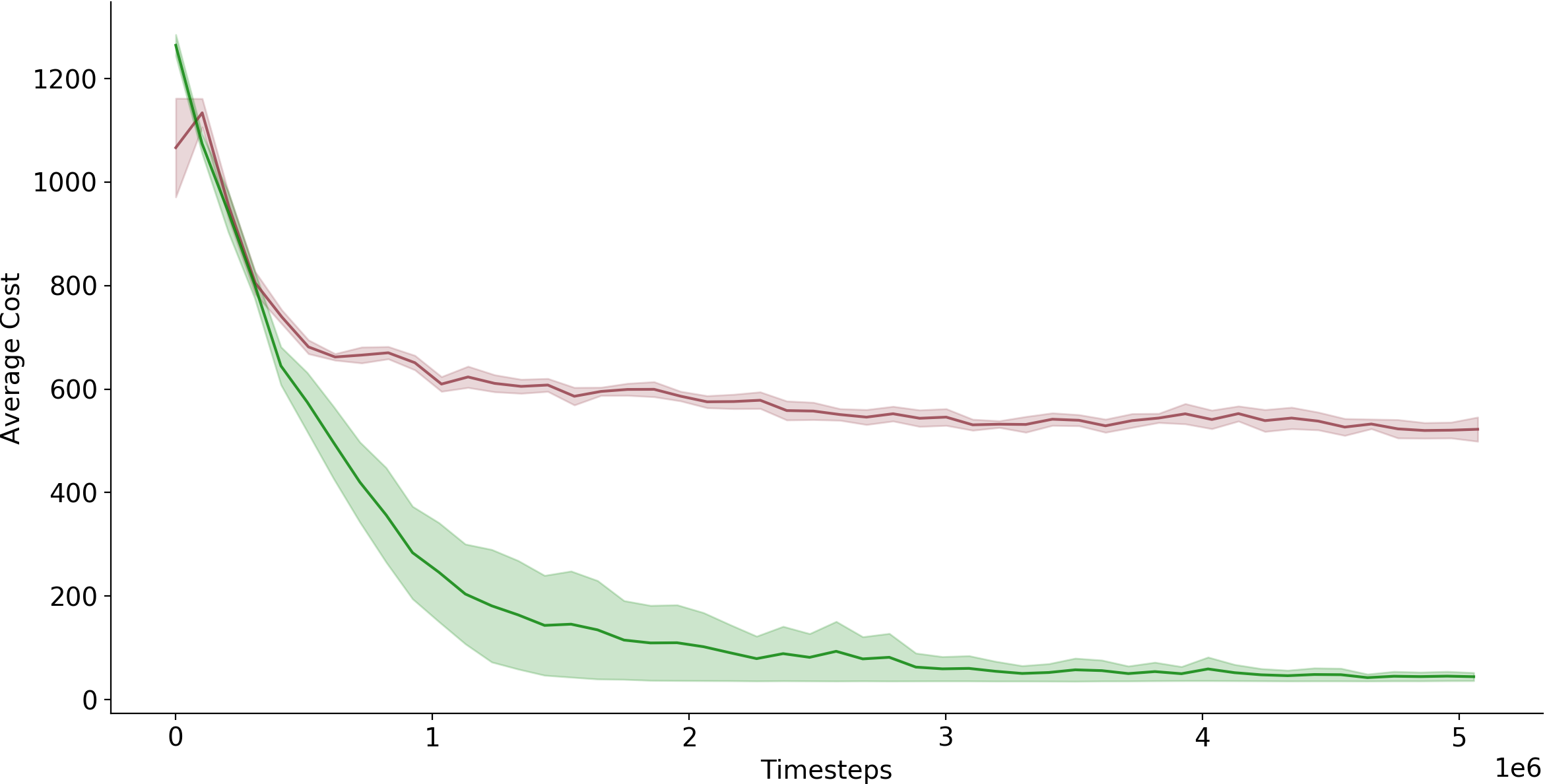} }\label{Fig:cost-wolf}}
\subfloat[Orc]
{{\includegraphics[width=0.33\linewidth]{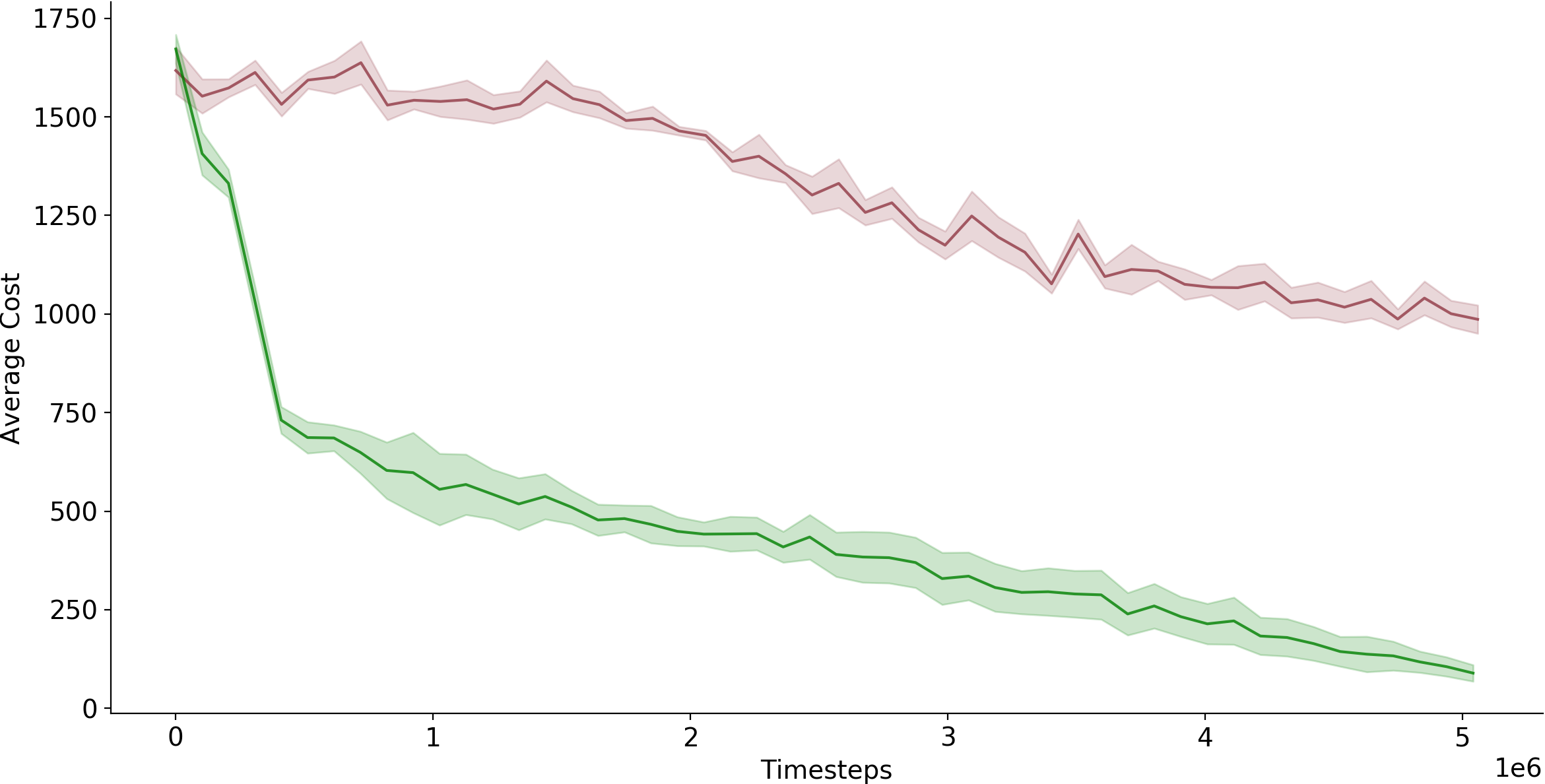} }\label{Fig:cost-orc}}
\subfloat[Mech]{{\includegraphics[width=0.33\linewidth]{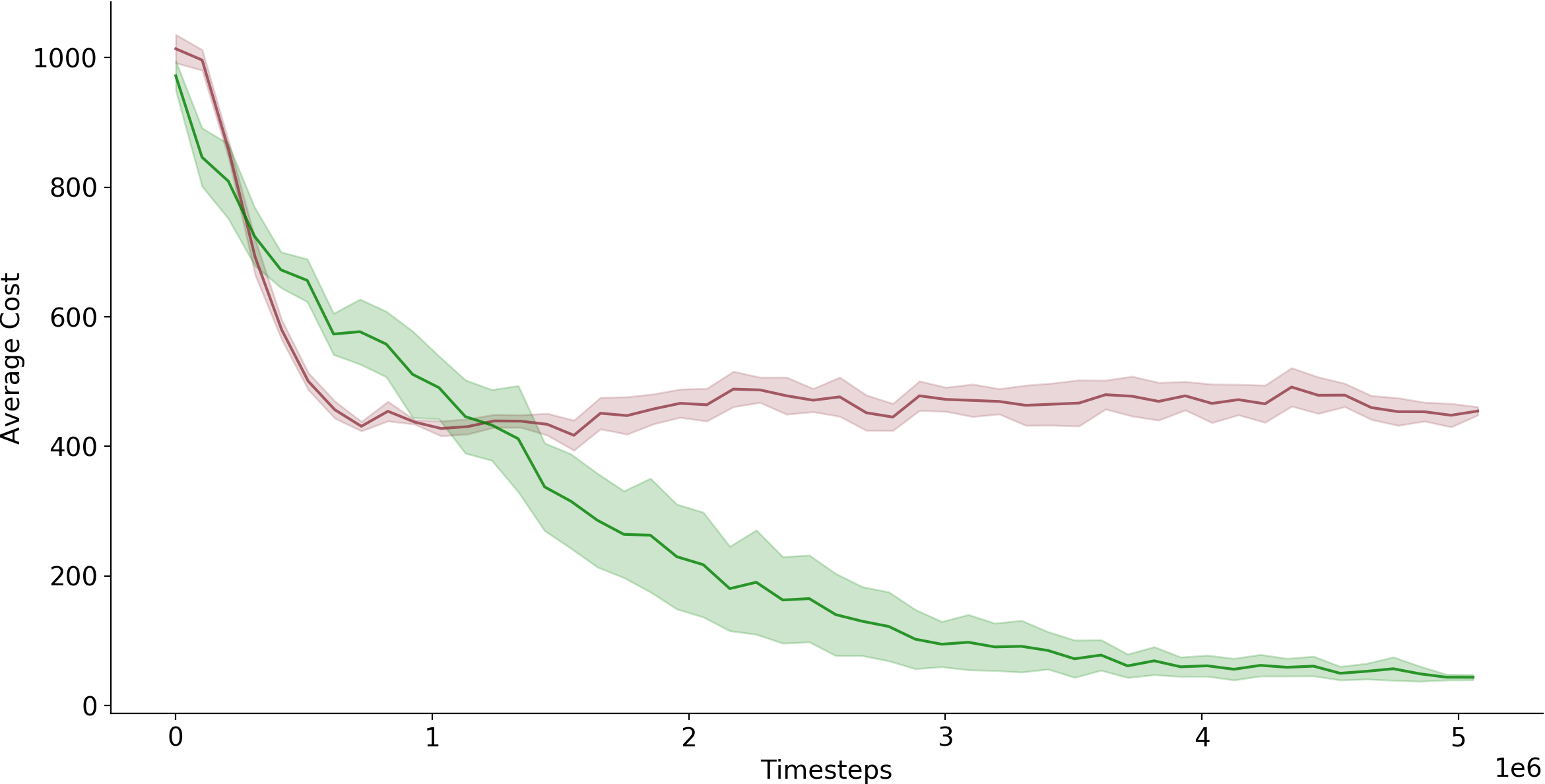} }\label{Fig:cost-mech}}
\caption{Comparing the performance, in terms of FDI-MCTS cost function, when using the DeepMimic-style reward function explained in Section \ref{subsec-reward-function} instead of directly using FDI-MCTS cost function as reward.}
\label{fig:cost_plot}
\end{figure*}

\subsubsection{FDI-MCTS is efficient in producing reference motions}
All reference motions were produced in less than five minutes. The initial and final frames of the cycles produced in the reference motion generation stage are pretty similar, although not exactly the same. However, slight differences are not a problem since in the self-imitation learning stage, the agent tries to imitate the reference cycle as much as the reward function allows it to. If the reference cycle is not perfect (which happens in most of the cases), the agent compensates the gap between initial and final frames such that the maximum reward in both frames is produced. This results in a motion that resembles the reference cycle as much as possible.

\subsubsection{Termination Curriculum improves training}
The plots in Fig. \ref{fig:reward_plot} show how different termination strategies work in terms of average reward. Compared to other versions, the termination curriculum shows superior performance for all characters except the wolf, where its performance is similar to the tight threshold strategy. The reason is that the wolf character, unlike orc and mech, is a quadruped and in the case of quadrupeds, a wide range of policies (including the random policies) can easily avoid the character from falling down. This results in receiving high rewards in Fig. \ref{Fig:reward-wolf}, and the curriculum has less effect. However, as it can be seen for the orc (Fig. \ref{Fig:reward-orc}) and mech (Fig. \ref{Fig:reward-mech}) characters, the tight threshold strategy shows instabilities since it uses a very optimistic reward threshold, which makes it fragile to small noise in the environment. On the other hand, the termination curriculum is producing stable results over all five runs for the three characters.

Another observation in Fig. \ref{fig:reward_plot} is how the termination curriculum significantly decreases the sample complexity when compared to the naive version with no termination. This, together with multi-threading, enabled us to successfully train the agents in only a few hours (including the reference motion generation stage). This is very promising since it can be used by game developers and animators as an easy-to-use and cheap animation production pipeline.

\subsubsection{The combination of FDI-MCTS and PPO is better than PPO alone}
Fig. \ref{fig:cost_plot} plots the FDI-MCTS cost when using PPO with DeepMimic-style reward function $r_t$, explained in Section \ref{subsec-reward-function}, as opposed to the reward function $r_t=\frac{b-\mathcal{J}_t}{b}$, which directly optimizes the FDI-MCTS objective function $\mathcal{J}_t$, while at the same time avoiding very large and small rewards, which is required for PPO's value function predictor network training. As it can be seen in the figure, the DeepMimic-style reward acts as a good proxy for optimizing the FDI-MCTS cost. On the other hand, trying to directly apply PPO to the locomotion problem without the FDI-MCTS imitation reward leads to clearly inferior results.

\section{Discussion and Conclusions}\label{Sec-Discussion}
We proposed an approach for constructing a policy network for synthesizing stable locomotion movements for arbitrary character anatomies. Our approach starts by running an online optimization method (FDI-MCTS \citep{Rajamaeki2018}) to generate a stable locomotion gait. It then extracts a cyclic motion out of the generated movement as the reference motion. In the next stage, inspired by DeepMimic \citep{2018-TOG-deepMimic}, proximal policy optimization (PPO) \citep{schulman2017proximal} is used such that the character is able to accomplish the locomotion task while imitating the reference motion. We also proposed Termination Curriculum (TC), a continuous curriculum learning mechanism to enable rapid training of the final policy. The core idea of this mechanism is to terminate the episode if the instantaneous reward becomes lower than some threshold $R^{min}_t$. Decreasing this threshold during the training results in a continuous curriculum which limits the state space regions that are visible to the agent during training.

Our experiments show that the proposed FDI-MCTS to PPO pipeline combines the advantages of both algorithms. FDI-MCTS allows rapid discovery and visualization of behaviors, enabling fast reward function design iteration. FDI-MCTS also provides more flexibility in the reward design, as one can simply use quadratic cost terms without worrying about excessive reward magnitude. Once a suitable gait has been found, PPO with DeepMimic reward function can produce a stable and computationally efficient neural network policy. In contrast, using FDI-MCTS alone incurs orders of magnitude higher runtime cost due to the forward simulation, and PPO alone prevents the fast reward design iteration. In absence of the FDI-MCTS -generated reference motion, PPO also failed in optimizing our locomotion reward function.

Although our approach improves the sample complexity of DeepMimic-style learning, it still has a high sample complexity. In future work, this could be improved by using more recent state-of-the-art reinforcement learning algorithms, such as Maximum a Posteriori Policy Optimisation (MPO) \citep{abdolmaleki2018maximum} or Soft Actor-Critic (SAC) \citep{haarnoja2018soft}. However, even with a more advanced RL algorithm, our proposed approach of using a trajectory optimization method for reference movement generation will probably offer faster reward function and movement style design iteration, as opposed to using RL alone.

\begin{acks}
This work has been supported by the Academy of Finland grants $299358$ and $305737$. The authors would like to thank the reviewers for their useful comments for improving this manuscript. Amin Babadi additionally thanks Joose Rajam\"{a}ki for his useful discussions and continuous support.
\end{acks}
\bibliographystyle{ACM-Reference-Format}
\bibliography{bib_source}


\begin{thebibliography}{30}


\ifx \showCODEN    \undefined \def \showCODEN     #1{\unskip}     \fi
\ifx \showDOI      \undefined \def \showDOI       #1{#1}\fi
\ifx \showISBNx    \undefined \def \showISBNx     #1{\unskip}     \fi
\ifx \showISBNxiii \undefined \def \showISBNxiii  #1{\unskip}     \fi
\ifx \showISSN     \undefined \def \showISSN      #1{\unskip}     \fi
\ifx \showLCCN     \undefined \def \showLCCN      #1{\unskip}     \fi
\ifx \shownote     \undefined \def \shownote      #1{#1}          \fi
\ifx \showarticletitle \undefined \def \showarticletitle #1{#1}   \fi
\ifx \showURL      \undefined \def \showURL       {\relax}        \fi
\providecommand\bibfield[2]{#2}
\providecommand\bibinfo[2]{#2}
\providecommand\natexlab[1]{#1}
\providecommand\showeprint[2][]{arXiv:#2}

\bibitem[\protect\citeauthoryear{Abadi, Agarwal, Barham, Brevdo, Chen, Citro,
  Corrado, Davis, Dean, Devin, Ghemawat, Goodfellow, Harp, Irving, Isard, Jia,
  Jozefowicz, Kaiser, Kudlur, Levenberg, Man\'{e}, Monga, Moore, Murray, Olah,
  Schuster, Shlens, Steiner, Sutskever, Talwar, Tucker, Vanhoucke, Vasudevan,
  Vi\'{e}gas, Vinyals, Warden, Wattenberg, Wicke, Yu, and Zheng}{Abadi
  et~al\mbox{.}}{2015}]%
        {tensorflow2015-whitepaper}
\bibfield{author}{\bibinfo{person}{Mart\'{\i}n Abadi}, \bibinfo{person}{Ashish
  Agarwal}, \bibinfo{person}{Paul Barham}, \bibinfo{person}{Eugene Brevdo},
  \bibinfo{person}{Zhifeng Chen}, \bibinfo{person}{Craig Citro},
  \bibinfo{person}{Greg~S. Corrado}, \bibinfo{person}{Andy Davis},
  \bibinfo{person}{Jeffrey Dean}, \bibinfo{person}{Matthieu Devin},
  \bibinfo{person}{Sanjay Ghemawat}, \bibinfo{person}{Ian Goodfellow},
  \bibinfo{person}{Andrew Harp}, \bibinfo{person}{Geoffrey Irving},
  \bibinfo{person}{Michael Isard}, \bibinfo{person}{Yangqing Jia},
  \bibinfo{person}{Rafal Jozefowicz}, \bibinfo{person}{Lukasz Kaiser},
  \bibinfo{person}{Manjunath Kudlur}, \bibinfo{person}{Josh Levenberg},
  \bibinfo{person}{Dandelion Man\'{e}}, \bibinfo{person}{Rajat Monga},
  \bibinfo{person}{Sherry Moore}, \bibinfo{person}{Derek Murray},
  \bibinfo{person}{Chris Olah}, \bibinfo{person}{Mike Schuster},
  \bibinfo{person}{Jonathon Shlens}, \bibinfo{person}{Benoit Steiner},
  \bibinfo{person}{Ilya Sutskever}, \bibinfo{person}{Kunal Talwar},
  \bibinfo{person}{Paul Tucker}, \bibinfo{person}{Vincent Vanhoucke},
  \bibinfo{person}{Vijay Vasudevan}, \bibinfo{person}{Fernanda Vi\'{e}gas},
  \bibinfo{person}{Oriol Vinyals}, \bibinfo{person}{Pete Warden},
  \bibinfo{person}{Martin Wattenberg}, \bibinfo{person}{Martin Wicke},
  \bibinfo{person}{Yuan Yu}, {and} \bibinfo{person}{Xiaoqiang Zheng}.}
  \bibinfo{year}{2015}\natexlab{}.
\newblock \bibinfo{title}{{TensorFlow}: Large-Scale Machine Learning on
  Heterogeneous Systems}.
\newblock
\newblock
\urldef\tempurl%
\url{https://www.tensorflow.org/}
\showURL{%
\tempurl}
\newblock
\shownote{Software available from tensorflow.org.}


\bibitem[\protect\citeauthoryear{Abdolmaleki, Springenberg, Tassa, Munos,
  Heess, and Riedmiller}{Abdolmaleki et~al\mbox{.}}{2018}]%
        {abdolmaleki2018maximum}
\bibfield{author}{\bibinfo{person}{Abbas Abdolmaleki},
  \bibinfo{person}{Jost~Tobias Springenberg}, \bibinfo{person}{Yuval Tassa},
  \bibinfo{person}{Remi Munos}, \bibinfo{person}{Nicolas Heess}, {and}
  \bibinfo{person}{Martin Riedmiller}.} \bibinfo{year}{2018}\natexlab{}.
\newblock \showarticletitle{Maximum a posteriori policy optimisation}.
\newblock \bibinfo{journal}{\emph{arXiv preprint arXiv:1806.06920}}
  (\bibinfo{year}{2018}).
\newblock


\bibitem[\protect\citeauthoryear{Babadi, Naderi, and
  H{\"a}m{\"a}l{\"a}inen}{Babadi et~al\mbox{.}}{2018}]%
        {babadi2018intelligent}
\bibfield{author}{\bibinfo{person}{Amin Babadi}, \bibinfo{person}{Kourosh
  Naderi}, {and} \bibinfo{person}{Perttu H{\"a}m{\"a}l{\"a}inen}.}
  \bibinfo{year}{2018}\natexlab{}.
\newblock \showarticletitle{Intelligent middle-level game control}. In
  \bibinfo{booktitle}{\emph{Proceedings of IEEE Conference on Computational
  Intelligence and Games (IEEE CIG)}}. IEEE.
\newblock


\bibitem[\protect\citeauthoryear{Bengio, Louradour, Collobert, and
  Weston}{Bengio et~al\mbox{.}}{2009}]%
        {bengio2009curriculum}
\bibfield{author}{\bibinfo{person}{Yoshua Bengio},
  \bibinfo{person}{J{\'e}r{\^o}me Louradour}, \bibinfo{person}{Ronan
  Collobert}, {and} \bibinfo{person}{Jason Weston}.}
  \bibinfo{year}{2009}\natexlab{}.
\newblock \showarticletitle{Curriculum learning}. In
  \bibinfo{booktitle}{\emph{Proceedings of International Conference on Machine
  Learning (ICML)}}. ACM, \bibinfo{pages}{41--48}.
\newblock


\bibitem[\protect\citeauthoryear{Browne, Powley, Whitehouse, Lucas, Cowling,
  Rohlfshagen, Tavener, Perez, Samothrakis, and Colton}{Browne
  et~al\mbox{.}}{2012}]%
        {browne2012survey}
\bibfield{author}{\bibinfo{person}{Cameron~B Browne}, \bibinfo{person}{Edward
  Powley}, \bibinfo{person}{Daniel Whitehouse}, \bibinfo{person}{Simon~M
  Lucas}, \bibinfo{person}{Peter~I Cowling}, \bibinfo{person}{Philipp
  Rohlfshagen}, \bibinfo{person}{Stephen Tavener}, \bibinfo{person}{Diego
  Perez}, \bibinfo{person}{Spyridon Samothrakis}, {and} \bibinfo{person}{Simon
  Colton}.} \bibinfo{year}{2012}\natexlab{}.
\newblock \showarticletitle{A survey of monte carlo tree search methods}.
\newblock \bibinfo{journal}{\emph{IEEE Transactions on Computational
  Intelligence and AI in games}} \bibinfo{volume}{4}, \bibinfo{number}{1}
  (\bibinfo{year}{2012}), \bibinfo{pages}{1--43}.
\newblock


\bibitem[\protect\citeauthoryear{Geijtenbeek and Pronost}{Geijtenbeek and
  Pronost}{2012}]%
        {geijtenbeek2012interactive}
\bibfield{author}{\bibinfo{person}{Thomas Geijtenbeek} {and}
  \bibinfo{person}{Nicolas Pronost}.} \bibinfo{year}{2012}\natexlab{}.
\newblock \showarticletitle{Interactive character animation using simulated
  physics: A state-of-the-art review}. In \bibinfo{booktitle}{\emph{Computer
  Graphics Forum}}, Vol.~\bibinfo{volume}{31}. Wiley Online Library,
  \bibinfo{pages}{2492--2515}.
\newblock


\bibitem[\protect\citeauthoryear{Geijtenbeek, van~de Panne, and van~der
  Stappen}{Geijtenbeek et~al\mbox{.}}{2013}]%
        {Geijtenbeek2013}
\bibfield{author}{\bibinfo{person}{Thomas Geijtenbeek},
  \bibinfo{person}{Michiel van~de Panne}, {and} \bibinfo{person}{A.~Frank
  van~der Stappen}.} \bibinfo{year}{2013}\natexlab{}.
\newblock \showarticletitle{Flexible muscle-based locomotion for bipedal
  creatures}.
\newblock \bibinfo{journal}{\emph{ACM Transactions on Graphics (TOG)}}
  \bibinfo{volume}{32}, \bibinfo{number}{6} (\bibinfo{year}{2013}).
\newblock


\bibitem[\protect\citeauthoryear{Haarnoja, Zhou, Abbeel, and Levine}{Haarnoja
  et~al\mbox{.}}{2018}]%
        {haarnoja2018soft}
\bibfield{author}{\bibinfo{person}{Tuomas Haarnoja}, \bibinfo{person}{Aurick
  Zhou}, \bibinfo{person}{Pieter Abbeel}, {and} \bibinfo{person}{Sergey
  Levine}.} \bibinfo{year}{2018}\natexlab{}.
\newblock \showarticletitle{Soft actor-critic: Off-policy maximum entropy deep
  reinforcement learning with a stochastic actor}.
\newblock \bibinfo{journal}{\emph{arXiv preprint arXiv:1801.01290}}
  (\bibinfo{year}{2018}).
\newblock


\bibitem[\protect\citeauthoryear{H{\"a}m{\"a}l{\"a}inen, Babadi, Ma, and
  Lehtinen}{H{\"a}m{\"a}l{\"a}inen et~al\mbox{.}}{2018}]%
        {hamalainen2018ppo}
\bibfield{author}{\bibinfo{person}{Perttu H{\"a}m{\"a}l{\"a}inen},
  \bibinfo{person}{Amin Babadi}, \bibinfo{person}{Xiaoxiao Ma}, {and}
  \bibinfo{person}{Jaakko Lehtinen}.} \bibinfo{year}{2018}\natexlab{}.
\newblock \showarticletitle{{P}{P}{O}-{C}{M}{A}: Proximal Policy Optimization
  with Covariance Matrix Adaptation}.
\newblock \bibinfo{journal}{\emph{arXiv preprint arXiv:1810.02541}}
  (\bibinfo{year}{2018}).
\newblock


\bibitem[\protect\citeauthoryear{H{\"a}m{\"a}l{\"a}inen, Eriksson, Tanskanen,
  Kyrki, and Lehtinen}{H{\"a}m{\"a}l{\"a}inen et~al\mbox{.}}{2014}]%
        {hamalainen2014online}
\bibfield{author}{\bibinfo{person}{Perttu H{\"a}m{\"a}l{\"a}inen},
  \bibinfo{person}{Sebastian Eriksson}, \bibinfo{person}{Esa Tanskanen},
  \bibinfo{person}{Ville Kyrki}, {and} \bibinfo{person}{Jaakko Lehtinen}.}
  \bibinfo{year}{2014}\natexlab{}.
\newblock \showarticletitle{Online motion synthesis using sequential monte
  carlo}.
\newblock \bibinfo{journal}{\emph{ACM Transactions on Graphics (TOG)}}
  \bibinfo{volume}{33}, \bibinfo{number}{4} (\bibinfo{year}{2014}),
  \bibinfo{pages}{51}.
\newblock


\bibitem[\protect\citeauthoryear{Hansen}{Hansen}{2006}]%
        {hansen2006cma}
\bibfield{author}{\bibinfo{person}{Nikolaus Hansen}.}
  \bibinfo{year}{2006}\natexlab{}.
\newblock \showarticletitle{The CMA evolution strategy: a comparing review}.
\newblock In \bibinfo{booktitle}{\emph{Towards a new evolutionary
  computation}}. \bibinfo{publisher}{Springer}, \bibinfo{pages}{75--102}.
\newblock


\bibitem[\protect\citeauthoryear{Liu and Hodgins}{Liu and Hodgins}{2018}]%
        {liu2018learning}
\bibfield{author}{\bibinfo{person}{Libin Liu} {and} \bibinfo{person}{Jessica
  Hodgins}.} \bibinfo{year}{2018}\natexlab{}.
\newblock \showarticletitle{Learning basketball dribbling skills using
  trajectory optimization and deep reinforcement learning}.
\newblock \bibinfo{journal}{\emph{ACM Transactions on Graphics (TOG)}}
  \bibinfo{volume}{37}, \bibinfo{number}{4} (\bibinfo{year}{2018}),
  \bibinfo{pages}{142}.
\newblock


\bibitem[\protect\citeauthoryear{Liu, Panne, and Yin}{Liu
  et~al\mbox{.}}{2016}]%
        {liu2016guided}
\bibfield{author}{\bibinfo{person}{Libin Liu}, \bibinfo{person}{Michiel Van~De
  Panne}, {and} \bibinfo{person}{KangKang Yin}.}
  \bibinfo{year}{2016}\natexlab{}.
\newblock \showarticletitle{Guided learning of control graphs for physics-based
  characters}.
\newblock \bibinfo{journal}{\emph{ACM Transactions on Graphics (TOG)}}
  \bibinfo{volume}{35}, \bibinfo{number}{3} (\bibinfo{year}{2016}),
  \bibinfo{pages}{29}.
\newblock


\bibitem[\protect\citeauthoryear{Mnih, Kavukcuoglu, Silver, Rusu, Veness,
  Bellemare, Graves, Riedmiller, Fidjeland, Ostrovski, et~al\mbox{.}}{Mnih
  et~al\mbox{.}}{2015}]%
        {mnih2015human}
\bibfield{author}{\bibinfo{person}{Volodymyr Mnih}, \bibinfo{person}{Koray
  Kavukcuoglu}, \bibinfo{person}{David Silver}, \bibinfo{person}{Andrei~A
  Rusu}, \bibinfo{person}{Joel Veness}, \bibinfo{person}{Marc~G Bellemare},
  \bibinfo{person}{Alex Graves}, \bibinfo{person}{Martin Riedmiller},
  \bibinfo{person}{Andreas~K Fidjeland}, \bibinfo{person}{Georg Ostrovski},
  {et~al\mbox{.}}} \bibinfo{year}{2015}\natexlab{}.
\newblock \showarticletitle{Human-level control through deep reinforcement
  learning}.
\newblock \bibinfo{journal}{\emph{Nature}} \bibinfo{volume}{518},
  \bibinfo{number}{7540} (\bibinfo{year}{2015}), \bibinfo{pages}{529}.
\newblock


\bibitem[\protect\citeauthoryear{Naderi, Babadi, and
  H{\"a}m{\"a}l{\"a}inen}{Naderi et~al\mbox{.}}{2018}]%
        {naderi2018learning}
\bibfield{author}{\bibinfo{person}{Kourosh Naderi}, \bibinfo{person}{Amin
  Babadi}, {and} \bibinfo{person}{Perttu H{\"a}m{\"a}l{\"a}inen}.}
  \bibinfo{year}{2018}\natexlab{}.
\newblock \showarticletitle{Learning Physically Based Humanoid Climbing
  Movements}. In \bibinfo{booktitle}{\emph{Computer Graphics Forum}},
  Vol.~\bibinfo{volume}{37}. Wiley Online Library, \bibinfo{pages}{69--80}.
\newblock


\bibitem[\protect\citeauthoryear{Naderi, Rajam\"{a}ki, and
  H\"{a}m\"{a}l\"{a}inen}{Naderi et~al\mbox{.}}{2017}]%
        {naderi2017discovering}
\bibfield{author}{\bibinfo{person}{Kourosh Naderi}, \bibinfo{person}{Joose
  Rajam\"{a}ki}, {and} \bibinfo{person}{Perttu H\"{a}m\"{a}l\"{a}inen}.}
  \bibinfo{year}{2017}\natexlab{}.
\newblock \showarticletitle{Discovering and synthesizing humanoid climbing
  movements}.
\newblock \bibinfo{journal}{\emph{ACM Transactions on Graphics (TOG)}}
  \bibinfo{volume}{36}, \bibinfo{number}{4}, Article \bibinfo{articleno}{43}
  (\bibinfo{date}{July} \bibinfo{year}{2017}), \bibinfo{numpages}{11}~pages.
\newblock
\showISSN{0730-0301}
\urldef\tempurl%
\url{https://doi.org/10.1145/3072959.3073707}
\showDOI{\tempurl}


\bibitem[\protect\citeauthoryear{Peng, Abbeel, Levine, and van~de Panne}{Peng
  et~al\mbox{.}}{2018a}]%
        {2018-TOG-deepMimic}
\bibfield{author}{\bibinfo{person}{Xue~Bin Peng}, \bibinfo{person}{Pieter
  Abbeel}, \bibinfo{person}{Sergey Levine}, {and} \bibinfo{person}{Michiel
  van~de Panne}.} \bibinfo{year}{2018}\natexlab{a}.
\newblock \showarticletitle{{DeepMimic: Example-guided deep reinforcement
  learning of physics-based character skills}}.
\newblock \bibinfo{journal}{\emph{ACM Transactions on Graphics (TOG)}}
  \bibinfo{volume}{37}, \bibinfo{number}{4}, Article \bibinfo{articleno}{143}
  (\bibinfo{date}{July} \bibinfo{year}{2018}), \bibinfo{numpages}{14}~pages.
\newblock
\showISSN{0730-0301}
\urldef\tempurl%
\url{https://doi.org/10.1145/3197517.3201311}
\showDOI{\tempurl}


\bibitem[\protect\citeauthoryear{Peng, Berseth, Yin, and Van De~Panne}{Peng
  et~al\mbox{.}}{2017}]%
        {peng2017deeploco}
\bibfield{author}{\bibinfo{person}{Xue~Bin Peng}, \bibinfo{person}{Glen
  Berseth}, \bibinfo{person}{KangKang Yin}, {and} \bibinfo{person}{Michiel Van
  De~Panne}.} \bibinfo{year}{2017}\natexlab{}.
\newblock \showarticletitle{Deeploco: dynamic locomotion skills using
  hierarchical deep reinforcement learning}.
\newblock \bibinfo{journal}{\emph{ACM Transactions on Graphics (TOG)}}
  \bibinfo{volume}{36}, \bibinfo{number}{4} (\bibinfo{year}{2017}),
  \bibinfo{pages}{41}.
\newblock


\bibitem[\protect\citeauthoryear{Peng, Kanazawa, Malik, Abbeel, and
  Levine}{Peng et~al\mbox{.}}{2018b}]%
        {2018-TOG-SFV}
\bibfield{author}{\bibinfo{person}{Xue~Bin Peng}, \bibinfo{person}{Angjoo
  Kanazawa}, \bibinfo{person}{Jitendra Malik}, \bibinfo{person}{Pieter Abbeel},
  {and} \bibinfo{person}{Sergey Levine}.} \bibinfo{year}{2018}\natexlab{b}.
\newblock \showarticletitle{SFV: Reinforcement learning of physical skills from
  videos}.
\newblock \bibinfo{journal}{\emph{ACM Transactions on Graphics (TOG)}}
  \bibinfo{volume}{37}, \bibinfo{number}{6}, Article \bibinfo{articleno}{178}
  (\bibinfo{date}{Nov.} \bibinfo{year}{2018}), \bibinfo{numpages}{14}~pages.
\newblock


\bibitem[\protect\citeauthoryear{Rajam{\"a}ki and
  H{\"a}m{\"a}l{\"a}inen}{Rajam{\"a}ki and H{\"a}m{\"a}l{\"a}inen}{2017}]%
        {rajamaki2017augmenting}
\bibfield{author}{\bibinfo{person}{Joose Rajam{\"a}ki} {and}
  \bibinfo{person}{Perttu H{\"a}m{\"a}l{\"a}inen}.}
  \bibinfo{year}{2017}\natexlab{}.
\newblock \showarticletitle{Augmenting sampling based controllers with machine
  learning}. In \bibinfo{booktitle}{\emph{Proceedings of the ACM
  SIGGRAPH/Eurographics Symposium on Computer Animation}}. ACM,
  \bibinfo{pages}{11}.
\newblock


\bibitem[\protect\citeauthoryear{Rajam{\"a}ki and
  H{\"a}m{\"a}l{\"a}inen}{Rajam{\"a}ki and H{\"a}m{\"a}l{\"a}inen}{2018}]%
        {Rajamaeki2018}
\bibfield{author}{\bibinfo{person}{Joose~Julius Rajam{\"a}ki} {and}
  \bibinfo{person}{Perttu H{\"a}m{\"a}l{\"a}inen}.}
  \bibinfo{year}{2018}\natexlab{}.
\newblock \showarticletitle{Continuous control monte carlo tree search informed
  by multiple experts}.
\newblock \bibinfo{journal}{\emph{IEEE transactions on visualization and
  computer graphics}} (\bibinfo{year}{2018}).
\newblock


\bibitem[\protect\citeauthoryear{Salimans, Ho, Chen, Sidor, and
  Sutskever}{Salimans et~al\mbox{.}}{2017}]%
        {salimans2017evolution}
\bibfield{author}{\bibinfo{person}{Tim Salimans}, \bibinfo{person}{Jonathan
  Ho}, \bibinfo{person}{Xi Chen}, \bibinfo{person}{Szymon Sidor}, {and}
  \bibinfo{person}{Ilya Sutskever}.} \bibinfo{year}{2017}\natexlab{}.
\newblock \showarticletitle{Evolution strategies as a scalable alternative to
  reinforcement learning}.
\newblock \bibinfo{journal}{\emph{arXiv preprint arXiv:1703.03864}}
  (\bibinfo{year}{2017}).
\newblock


\bibitem[\protect\citeauthoryear{Schulman, Levine, Abbeel, Jordan, and
  Moritz}{Schulman et~al\mbox{.}}{2015}]%
        {schulman2015trust}
\bibfield{author}{\bibinfo{person}{John Schulman}, \bibinfo{person}{Sergey
  Levine}, \bibinfo{person}{Pieter Abbeel}, \bibinfo{person}{Michael Jordan},
  {and} \bibinfo{person}{Philipp Moritz}.} \bibinfo{year}{2015}\natexlab{}.
\newblock \showarticletitle{Trust region policy optimization}. In
  \bibinfo{booktitle}{\emph{Proceedings of International Conference on Machine
  Learning (ICML)}}. \bibinfo{pages}{1889--1897}.
\newblock


\bibitem[\protect\citeauthoryear{Schulman, Wolski, Dhariwal, Radford, and
  Klimov}{Schulman et~al\mbox{.}}{2017}]%
        {schulman2017proximal}
\bibfield{author}{\bibinfo{person}{John Schulman}, \bibinfo{person}{Filip
  Wolski}, \bibinfo{person}{Prafulla Dhariwal}, \bibinfo{person}{Alec Radford},
  {and} \bibinfo{person}{Oleg Klimov}.} \bibinfo{year}{2017}\natexlab{}.
\newblock \showarticletitle{Proximal policy optimization algorithms}.
\newblock \bibinfo{journal}{\emph{arXiv preprint arXiv:1707.06347}}
  (\bibinfo{year}{2017}).
\newblock


\bibitem[\protect\citeauthoryear{Silver, Huang, Maddison, Guez, Sifre, Van
  Den~Driessche, Schrittwieser, Antonoglou, Panneershelvam, Lanctot,
  et~al\mbox{.}}{Silver et~al\mbox{.}}{2016}]%
        {silver2016mastering}
\bibfield{author}{\bibinfo{person}{David Silver}, \bibinfo{person}{Aja Huang},
  \bibinfo{person}{Chris~J Maddison}, \bibinfo{person}{Arthur Guez},
  \bibinfo{person}{Laurent Sifre}, \bibinfo{person}{George Van Den~Driessche},
  \bibinfo{person}{Julian Schrittwieser}, \bibinfo{person}{Ioannis Antonoglou},
  \bibinfo{person}{Veda Panneershelvam}, \bibinfo{person}{Marc Lanctot},
  {et~al\mbox{.}}} \bibinfo{year}{2016}\natexlab{}.
\newblock \showarticletitle{Mastering the game of Go with deep neural networks
  and tree search}.
\newblock \bibinfo{journal}{\emph{Nature}} \bibinfo{volume}{529},
  \bibinfo{number}{7587} (\bibinfo{year}{2016}), \bibinfo{pages}{484--489}.
\newblock


\bibitem[\protect\citeauthoryear{Silver, Schrittwieser, Simonyan, Antonoglou,
  Huang, Guez, Hubert, Baker, Lai, Bolton, et~al\mbox{.}}{Silver
  et~al\mbox{.}}{2017}]%
        {silver2017mastering}
\bibfield{author}{\bibinfo{person}{David Silver}, \bibinfo{person}{Julian
  Schrittwieser}, \bibinfo{person}{Karen Simonyan}, \bibinfo{person}{Ioannis
  Antonoglou}, \bibinfo{person}{Aja Huang}, \bibinfo{person}{Arthur Guez},
  \bibinfo{person}{Thomas Hubert}, \bibinfo{person}{Lucas Baker},
  \bibinfo{person}{Matthew Lai}, \bibinfo{person}{Adrian Bolton},
  {et~al\mbox{.}}} \bibinfo{year}{2017}\natexlab{}.
\newblock \showarticletitle{Mastering the game of go without human knowledge}.
\newblock \bibinfo{journal}{\emph{Nature}} \bibinfo{volume}{550},
  \bibinfo{number}{7676} (\bibinfo{year}{2017}), \bibinfo{pages}{354}.
\newblock


\bibitem[\protect\citeauthoryear{Sironi, Liu, Perez-Liebana, Gaina, Bravi,
  Lucas, and Winands}{Sironi et~al\mbox{.}}{2018}]%
        {sironi2018self}
\bibfield{author}{\bibinfo{person}{Chiara~F Sironi}, \bibinfo{person}{Jialin
  Liu}, \bibinfo{person}{Diego Perez-Liebana}, \bibinfo{person}{Raluca~D
  Gaina}, \bibinfo{person}{Ivan Bravi}, \bibinfo{person}{Simon~M Lucas}, {and}
  \bibinfo{person}{Mark~HM Winands}.} \bibinfo{year}{2018}\natexlab{}.
\newblock \showarticletitle{Self-adaptive mcts for general video game playing}.
  In \bibinfo{booktitle}{\emph{International Conference on the Applications of
  Evolutionary Computation}}. Springer, \bibinfo{pages}{358--375}.
\newblock


\bibitem[\protect\citeauthoryear{Smith}{Smith}{2001}]%
        {ode}
\bibfield{author}{\bibinfo{person}{Russell~L. Smith}.}
  \bibinfo{year}{2001}\natexlab{}.
\newblock \bibinfo{title}{Open Dynamics Engine}.
\newblock \bibinfo{howpublished}{\url{http://www.ode.org/}}.
\newblock
\newblock
\shownote{Accessed: 2019-01-01.}


\bibitem[\protect\citeauthoryear{Sutton and Barto}{Sutton and Barto}{2018}]%
        {sutton2018reinforcement}
\bibfield{author}{\bibinfo{person}{Richard~S Sutton} {and}
  \bibinfo{person}{Andrew~G Barto}.} \bibinfo{year}{2018}\natexlab{}.
\newblock \bibinfo{booktitle}{\emph{Reinforcement learning: An introduction}}.
\newblock \bibinfo{publisher}{MIT press}.
\newblock


\bibitem[\protect\citeauthoryear{Yu, Turk, and Liu}{Yu et~al\mbox{.}}{2018}]%
        {yu2018learning}
\bibfield{author}{\bibinfo{person}{Wenhao Yu}, \bibinfo{person}{Greg Turk},
  {and} \bibinfo{person}{C~Karen Liu}.} \bibinfo{year}{2018}\natexlab{}.
\newblock \showarticletitle{Learning symmetric and low-energy locomotion}.
\newblock \bibinfo{journal}{\emph{ACM Transactions on Graphics (TOG)}}
  \bibinfo{volume}{37}, \bibinfo{number}{4} (\bibinfo{year}{2018}),
  \bibinfo{pages}{144}.
\newblock


\end{thebibliography}


\end{document}